# Design and analytically full-wave validation of the invisibility cloaks, concentrators, and field rotators created with a general class of transformations

Yu Luo [1], Hongsheng Chen [1,2], Jingjing Zhang [1], Lixin Ran [1]*, and Jin Au Kong [1,2]

[1] *The Electromagnetics Academy at Zhejiang University, Zhejiang University, Hangzhou 310058, P. R. China*

[2] *Research Laboratory of Electronics, Massachusetts Institute of Technology, Cambridge, Massachusetts 02139*

**Abstract**

We investigate a general class of electromagnetic devices created with any continuous transformation functions by rigorously calculating the analytical expressions of the electromagnetic field in the whole space. Some interesting phenomena associated with these transformation devices, including the invisibility cloaks, concentrators, and field rotators, are discussed. By carefully choosing the transformation function, we can realize cloaks which are insensitive to perturbations at both the inner and outer boundaries. Furthermore, we find that when the coating layer of the concentrator is realized with left-handed materials, energy will circulate between the coating and the core, and the energy transmits through the core of the concentrator can be much bigger than that transmits through the concentrator. Therefore, such concentrator is also a power flux amplifier. Finally, we propose a spherical field rotator, which functions as not only a wave vector rotator, but also a polarization rotator, depending on the orientations of the spherical rotator with respect to the incident wave direction. The functionality of these novel transformation devices are all successfully confirmed by our analytical full wave method, which also provides an alternate computational efficient validation method in contrast to numerical validation methods.

*\* Author to whom correspondence should be addressed; electronic mail: ranlx@zju.edu.cn*

## I. INTRODUCTION

Recently, cloaks of invisibility have received much attention [1-17]. Pendry *et al.* firstly proposed a coordinate transformation approach to provide a new method to control EM fields, by which a space consisting of the normal free space can be squeezed into a new space with different volume and space-distributed constitutive parameters [1]. Following this approach, a microwave invisibility cloak was soon proposed and experimentally realized [2], and some other novel devices, such as the EM concentrators [3, 4], rotators [5], and hyperlens [6, 7] were also investigated by similar methods. Apart from the pure transformation method, full wave simulations [8] and traditional EM analysis based on Mie scattering [9] were also presented and both verified the validation of the transformation approach. These rapid progresses make the coordinate transformation approach be a hot topic in the electromagnetics community and imply very important future applications [10, 11]. The current discussions on the invisibility cloak are mostly based on a linear transformation presented by Pendry *et al.*[1]. Some authors have considered invisibility cloak with high order transformations [12, 13]. In this paper, we try to present a generalized formulation on how to do the transformations to obtain different devices by combining the merits of both coordinate transformation and the Mie scattering solutions. Starting from such a formulation, aforementioned devices (invisibility cloaks, concentrators, and field rotators) can be easily obtained by simply choosing different scalar transformation functions (including the transformation on $\theta$ and $\phi$), and the internal field distributions in these devices can also be controlled by tuning the shapes of the functions. We show that such scalar transformation functions can be any continuous functions, and can be applied to any coordinate to realize electromagnetic devices with novel functionality. This analytical full wave method provides not only a global physical understanding to the effect of the coordinate transformation, but also a convenient analysis and design tool for such novel devices due to its very high computational efficiency.

## II. FOMULATIONS

Consider the three dimensional case (the following idea is also applicable to two dimensional cylindrical case): a general coordinate transformation between two spherical coordinate systems ($r', \theta', \varphi'$) and ($r, \theta, \varphi$) is described by

$$r' = f(r,\theta,\varphi), \quad \theta' = g(r,\theta,\varphi), \quad \varphi' = h(r,\theta,\varphi), \tag{1}$$

where $r', \theta', \varphi'$ represent the coordinates in the original coordinate system and $f(\bullet)$, $g(\bullet)$, $h(\bullet)$ can be arbitrary continuous functions. Following the transformation approach proposed in [1], Maxwell equations still remain its form invariance in the new space ($r, \theta, \varphi$) but the permittivity and permeability will turn into distributed, or space dependent tensors, i.e.,

$$\bar{\bar{\varepsilon}} = \varepsilon_0 \bar{\bar{T}}^{-1}, \quad \bar{\bar{\mu}} = \mu_0 \bar{\bar{T}}^{-1}, \tag{2}$$

where $\varepsilon_0$ and $\mu_0$ represent the scalar permittivity and permeability of free space in the original space before transformation. The matrix $T$ is defined by $T = \bar{\bar{J}}^T \bar{\bar{J}} / \det(\bar{\bar{J}})$, where $\bar{\bar{J}} = \dfrac{\partial(f,g,h)}{\partial(r,\theta,\varphi)}$ is the Jacobian matrix [14, 15]. According to the Mie scattering theory, for source free cases, we can decompose the fields into TE and TM modes by introducing the vector potential $\bar{A}_{TE}$ and $\bar{A}_{TM}$ in the new space and express the fields as

$$\begin{cases} B_{TM} = \nabla \times (\bar{A}_{TM}) \\ D_{TM} = \dfrac{i}{\omega}\left[\nabla \times \left(\bar{\bar{\mu}}^{-1} \cdot \nabla \times (\bar{A}_{TM})\right)\right] \\ D_{TE} = -\nabla \times (\bar{A}_{TE}) \\ B_{TE} = \dfrac{i}{\omega}\left[\nabla \times \left(\bar{\bar{\varepsilon}}^{-1} \cdot \nabla \times (\bar{A}_{TE})\right)\right] \end{cases}, \quad (3)$$

where $B$ and $D$ represent the magnetic flux density and electric displacement, respectively. Since the media described by $\bar{\bar{\varepsilon}}$ and $\bar{\bar{\mu}}$ is no longer isotropic, the directions of $\bar{A}_{TE}$ and $\bar{A}_{TM}$ will not always be along the $r$ direction. For mathematical convenience, we let

$$\bar{A}_{TM} = \left(\dfrac{\partial f}{\partial r}\hat{r} + \dfrac{\partial f}{r\partial\theta}\hat{\theta} + \dfrac{\partial f}{r\sin\theta\partial\varphi}\hat{\varphi}\right)\Phi_{TM},$$
$$\bar{A}_{TE} = \left(\dfrac{\partial f}{\partial r}\hat{r} + \dfrac{\partial f}{r\partial\theta}\hat{\theta} + \dfrac{\partial f}{r\sin\theta\partial\varphi}\hat{\varphi}\right)\Phi_{TE}, \quad (4)$$

where $\Phi_{TM}$ and $\Phi_{TE}$ are scalar potentials for TE and TM cases, respectively. Note that if $f(\bullet)$ is only a function of $r$, for example, a linear function like $f(r) = \dfrac{R_2}{(R_2 - R_1)}(r - R_1)$, then the two vector potential $\bar{A}_{TE}$ and $\bar{A}_{TM}$ will be along the $r$ direction, and it will be reduced to the case studied in Ref [7]. Substituting equation (4) into equation (3), we obtain the partial differential equation for $\Phi_{TE}$ and $\Phi_{TM}$:

$$\left[\dfrac{\partial^2}{\partial f^2} + \dfrac{1}{f^2 \sin g}\dfrac{\partial}{\partial g}\left(\sin g \dfrac{\partial}{\partial g}\right) + \dfrac{1}{f^2 \sin^2 g}\dfrac{\partial^2}{\partial h^2} + k_0^2\right]\Phi = 0, \quad (5)$$

which takes the same form as the Helmholtz equation, so one of its special solution is

$$\Phi = \hat{B}_n(k_0 f) P_n^m(\cos g)(A_m \cos mh + B_m \sin mh), \quad (6)$$

where $\hat{B}_n(\xi)$ is Riccati-Bessel function, $P_n^m$ is the $n$ th orders of the associated

Legendre polynomials of degree $m$, and $A_m$ and $B_m$ are undetermined coefficients.

Using equation (4), we can obtain the vector potentials $\bar{A}_{TE}$ and $\bar{A}_{TM}$ as follows:

$$\bar{A}_{TM} = \sum_{m,n} a_{m,n}^{TM} \left( \frac{\partial f}{\partial r} \hat{r} + \frac{\partial f}{r \partial \theta} \hat{\theta} + \frac{\partial f}{r \sin\theta \partial \varphi} \hat{\varphi} \right) \hat{B}_n(k_0 f) P_n^m(\cos g)(A_m \cos mh + B_m \sin mh)$$

$$\bar{A}_{TE} = \sum_{m,n} a_{m,n}^{TE} \left( \frac{\partial f}{\partial r} \hat{r} + \frac{\partial f}{r \partial \theta} \hat{\theta} + \frac{\partial f}{r \sin\theta \partial \varphi} \hat{\varphi} \right) \hat{B}_n(k_0 f) P_n^m(\cos g)(A_m \cos mh + B_m \sin mh)$$

(7)

where the coefficients $a_{m,n}^{TM}$ and $a_{m,n}^{TE}$ can be determined by applying corresponding boundary conditions. Thus all the components of the total fields can be obtained by substituting equation (7) into equations (2), and take the following forms

$$E_r = \frac{i}{\omega \mu_0 \varepsilon_0} \left[ \frac{\partial f}{\partial r} \left( \frac{\partial^2}{\partial f^2} + k_0^2 \right) + \frac{\partial g}{\partial r} \frac{\partial^2}{\partial f \partial g} + \frac{\partial h}{\partial r} \frac{\partial^2}{\partial f \partial h} \right] \Phi_{TM}$$
$$+ \frac{1}{\varepsilon_0} \left( \sin g \frac{\partial h}{\partial r} \frac{\partial}{\partial g} - \frac{1}{\sin g} \frac{\partial g}{\partial r} \frac{\partial}{\partial h} \right) \Phi_{TE}$$

(8-1)

$$E_\theta = \frac{i}{\omega \mu_0 \varepsilon_0 r} \left[ \frac{\partial f}{\partial \theta} \left( \frac{\partial^2}{\partial f^2} + k_0^2 \right) + \frac{\partial g}{\partial \theta} \frac{\partial^2}{\partial f \partial g} + \frac{\partial h}{\partial \theta} \frac{\partial^2}{\partial f \partial h} \right] \Phi_{TM}$$
$$+ \frac{1}{\varepsilon_0 r} \left( \sin g \frac{\partial h}{\partial \theta} \frac{\partial}{\partial g} - \frac{1}{\sin g} \frac{\partial g}{\partial \theta} \frac{\partial}{\partial h} \right) \Phi_{TE}$$

(8-2)

$$E_\varphi = \frac{i}{\omega \mu_0 \varepsilon_0 r \sin\theta} \left[ \frac{\partial f}{\partial \varphi} \left( \frac{\partial^2}{\partial f^2} + k_0^2 \right) + \frac{\partial g}{\partial \varphi} \frac{\partial^2}{\partial f \partial g} + \frac{\partial h}{\partial \varphi} \frac{\partial^2}{\partial f \partial h} \right] \Phi_{TM}$$
$$+ \frac{1}{\varepsilon_0 r \sin\theta} \left( \sin g \frac{\partial h}{\partial \varphi} \frac{\partial}{\partial g} - \frac{1}{\sin g} \frac{\partial g}{\partial \varphi} \frac{\partial}{\partial h} \right) \Phi_{TE}$$

(8-3)

$$H_r = \frac{1}{\mu_0} \left( \frac{1}{\sin g} \frac{\partial g}{\partial r} \frac{\partial}{\partial h} - \sin g \frac{\partial h}{\partial r} \frac{\partial}{\partial g} \right) \Phi_{TM}$$
$$+ \frac{i}{\omega \mu_0 \varepsilon_0} \left[ \frac{\partial f}{\partial r} \left( \frac{\partial^2}{\partial f^2} + k_0^2 \right) + \frac{\partial g}{\partial r} \frac{\partial^2}{\partial f \partial g} + \frac{\partial h}{\partial r} \frac{\partial^2}{\partial f \partial h} \right] \Phi_{TE}$$

(8-4)

$$H_\theta = \frac{1}{\mu_0 r} \left( \frac{1}{\sin g} \frac{\partial g}{\partial \theta} \frac{\partial}{\partial h} - \sin g \frac{\partial h}{\partial \theta} \frac{\partial}{\partial g} \right) \Phi_{TM}$$
$$+ \frac{i}{\omega \mu_0 \varepsilon_0 r} \left[ \frac{\partial f}{\partial \theta} \left( \frac{\partial^2}{\partial f^2} + k_0^2 \right) + \frac{\partial g}{\partial \theta} \frac{\partial^2}{\partial f \partial g} + \frac{\partial h}{\partial \theta} \frac{\partial^2}{\partial f \partial h} \right] \Phi_{TE}$$

(8-5)

$$H_\varphi = \frac{1}{\mu_0 r \sin\theta}\left(\sin g \frac{\partial h}{\partial \varphi}\frac{\partial}{\partial g} - \frac{1}{\sin g}\frac{\partial g}{\partial \varphi}\frac{\partial}{\partial h}\right)\Phi_{TM}$$
$$+ \frac{i}{\omega\mu_0\varepsilon_0 r \sin\theta}\left[\frac{\partial f}{\partial \varphi}\left(\frac{\partial^2}{\partial f^2}+k_0^2\right)+\frac{\partial g}{\partial \varphi}\frac{\partial^2}{\partial f \partial g}+\frac{\partial h}{\partial \varphi}\frac{\partial^2}{\partial f \partial h}\right]\Phi_{TE} \quad (8\text{-}6)$$

where $E$ and $H$ represent the electric and magnetic fields, respectively.

### III. SPHERICAL CLOAKS

The above formulas are firstly applied to the spherical cloaks. Any continuous function $f(\bullet)$, $g(\bullet)$, $h(\bullet)$ that satisfy $f(R_2,\theta,\varphi)=R_2$, $g(R_2,\theta,\varphi)=\theta$, $h(R_2,\theta,\varphi)=\varphi+\varphi_0$ (where $\varphi_0$ is a definite constant) and $f(R_1,\theta,\varphi)=0$ (these conditions can be directly obtained from the partial differential equations by setting the scattering coefficients $T_n^{TM}$ and $T_n^{TE}$ to be zero at the outer boundary) can be used to achieve a coating the field inside which is always matched with the outer free space at the outer boundary $R_2$ and the potential is uniform everywhere at the inner boundary $R_1$. The coating with this quality can realize a perfect spherical invisibility cloak. Detailed illustration on it is out of the scope of this article and further discussion, with general theoretical analysis and numerical simulations will be given in our other paper. Here we only consider one simple case when $r'=f(r)$, $\theta'=\theta$, $\varphi'=\varphi$. The associated permittivity and permeability tensors are then given by:

$$\bar{\bar{\varepsilon}} = \varepsilon_r(r)\hat{r}\hat{r} + \varepsilon_t(r)\hat{\theta}\hat{\theta} + \varepsilon_t(r)\hat{\varphi}\hat{\varphi}, \qquad \bar{\bar{\mu}} = \mu_r(r)\hat{r}\hat{r} + \mu_t(r)\hat{\theta}\hat{\theta} + \mu_t(r)\hat{\varphi}\hat{\varphi},$$

where $\varepsilon_t = \varepsilon_0 f'(r)$, $\varepsilon_r = \varepsilon_0 \frac{f^2(r)}{r^2 f'(r)}$, $\mu_t = \mu_0 f'(r)$, and $\mu_r = \mu_0 \frac{f^2(r)}{r^2 f'(r)}$ (9)

For an arbitrary transformation function $f(r)$, we will show in detail that these parameters yield a perfect invisibility as long as $f(R_2)=R_2$ and $f(R_1)=0$ are satisfied.

Suppose an $E_x$ polarized plane wave with a unit amplitude $E_i = \hat{x}e^{ik_0z}$ is incident upon the coated sphere along the $z$ direction. With the solution of equation (4), the vector potentials for the incident fields ($r > R_2$), the scattered fields ($r > R_2$), and the fields inside the cloak layer ($R_1 < r < R_2$) can be written in the following forms respectively:

$$\bar{A}^i_{TM} = \hat{r}\frac{\cos\varphi}{\omega}\sum_n a_n\psi_n(k_0r)P^1_n(\cos\theta)$$
$$\bar{A}^i_{TE} = \hat{r}\frac{\sin\varphi}{\omega\eta_0}\sum_n a_n\psi_n(k_0r)P^1_n(\cos\theta)$$
(10-1)

$$\bar{A}^s_{TM} = \hat{r}\frac{\cos\varphi}{\omega}\sum_n a_n T_n^{TM}\zeta_n(k_0r)P^1_n(\cos\theta)$$
$$\bar{A}^s_{TE} = \hat{r}\frac{\sin\varphi}{\omega\eta_0}\sum_n a_n T_n^{TE}\zeta_n(k_0r)P^1_n(\cos\theta)$$
(10-2)

$$\bar{A}^c_{TM} = \hat{r}\frac{\cos\varphi}{\omega}\sum_n f'(r)\left(d_n^{TM}\psi_n(k_0f(r)) + f_n^{TM}\chi_n(k_0f(r))\right)P^1_n(\cos\theta)$$
$$\bar{A}^c_{TE} = \hat{r}\frac{\sin\varphi}{\omega\eta_0}\sum_n f'(r)\left(d_n^{TE}\psi_n(k_0f(r)) + f_n^{TE}\chi_n(k_0f(r))\right)P^1_n(\cos\theta)$$
(10-3)

where $a_n = \dfrac{(-i)^{-n}(2n+1)}{n(n+1)}$, $n=1,2,3,\ldots$, $\eta_0 = \sqrt{\dfrac{\mu_0}{\varepsilon_0}}$; $T_n^{TM}$, $T_n^{TE}$, $d_n^{TM}$, $d_n^{TE}$, $f_n^{TM}$ and $f_n^{TE}$ are unknown expansion coefficients; $\psi_n(\xi)$, $\chi_n(\xi)$, $\zeta_n(\xi)$ represent the Riccati-Bessel function of the first, the second, and the third kind, respectively [18]. Since $f(R_1)=0$, $\chi_n(0)$ is infinite, the finitude of the field at the inner boundary $R_1$ requires that $f_n^{TM} = f_n^{TE} = 0$ [9]. By applying the boundary conditions at the boundary of $r = R_2$, we can get other unknown coefficients:

$$T_n^{TM} = T_n^{TE} = -\frac{\psi'_n(k_0R_2)\psi_n(k_0f(R_2)) - \psi_n(k_0R_2)\psi'_n(k_0f(R_2))}{\zeta'_n(k_0R_2)\psi_n(k_0f(R_2)) - \zeta_n(k_0R_2)\psi'_n(k_0f(R_2))},$$
(11-1)

$$d_n^{TM} = d_n^{TE} = \frac{ia_n}{\zeta_n'(k_0 R_2)\psi_n(k_0 f(R_2)) - \zeta_n(k_0 R_2)\psi_n'(k_0 f(R_2))}, \quad (11\text{-}2)$$

Since $f(R_2) = R_2$, the above equations can be simplified as

$$T_n^{TM} = T_n^{TE} = 0, \quad d_n^{TM} = d_n^{TE} = 1, \quad (12)$$

The fact that coefficients $T_n^{TM}$ and $T_n^{TE}$ are exactly equal to zero indicates a reflectionless behavior of a perfect cloak. Substituting equations (10) into equations (8), after some algebraic manipulations, the summation $\sum_n$ can be written in closed forms. As a result, all components of the electric field are expressed as (Note that the parameters for free space can be regarded as $f(r) = r$, therefore the fields can still be written in the following forms):

$$E_r = f'(r)\sin\theta\cos\varphi e^{ik_0 f(r)\cos\theta}, \quad (13\text{-}1)$$

$$E_\theta = \frac{f(r)}{r}\cos\theta\cos\varphi e^{ik_0 f(r)\cos\theta}, \quad (13\text{-}2)$$

$$E_\varphi = -\frac{f(r)}{r}\sin\varphi e^{ik_0 f(r)\cos\theta}, \quad (13\text{-}3)$$

Therefore, from equation (12) and (13), we confirmed that as long as $f(R_2) = R_2$ and $f(R_1) = 0$, any spherical shell with parameters defined by equation (9) can yield a perfect invisibility. Different $f(r)$ in the region $R_1 < r < R_2$ will only cause different field distribution in the cloak layer, but will not disturb the field outside.

The distribution of the field in the cloak shell $R_1 < r < R_2$ and the sensitivity of the cloak to the perturbations at the boundary are determined by the transformation function $f(r)$. We investigate four types of cloak created with four different transformation functions: (Case I) $f_1(r) = \frac{R_2}{(R_2 - R_1)}(r - R_1)$, (Case II)

$f_2(r) = \frac{R_2}{(R_2 - R_1)^2}(r - R_1)^2$ with $f_2'(R_1) = 0$, (Case III) $f_3(r) = -\frac{R_2}{(R_2 - R_1)^2}(r - R_2)^2 + R_2$ with $f_3'(R_2) = 0$, and (Case IV) $f_4(r)$ with $f_4'(R_1) = 0$ and $f_4'(R_2) = 0$. Fig. 1 (a) displays the curves of the four transformation functions. Fig. 1 (b) shows the corresponding tangential and radial components of $\varepsilon$ and $\mu$ of the four different cloaks calculated from Equation (9). Fig. 1 (c), (d), (e), (f) depict the calculated $E_x$ fields distributions and Poynting vectors due to $E_x$ polarized wave incidence onto these four different cloaks, respectively. All the cloaks have a same size of $R_1 = 0.1$ m, and $R_2 = 0.2$ m. The wavelength in free space is 0.15 m. All the quantities are normalized to unity in this and the following calculations.

With different transformations, the fields inside the cloaks are differently distributed while the wave propagating in the outer region of the cloak remains undisturbed. In Fig.1 (c), the field is nearly uniformly distributed in the cloak shell with linear function $f_1(r) = \frac{R_2}{(R_2 - R_1)}(r - R_1)$ (transformation function used in Ref [1]) between $R_1$ and $R_2$. From the result we can find that this kind of cloak, which can be called linear-transformed cloak, is sensitive to perturbations both at the inner boundary $R_1$ and outer boundary $R_2$; In Fig.1 (d), the field is mainly distributed near the outer boundary in the cloak with the convex transformation function $f_2(r)$. And this so called convex-transformed cloak is not sensitive to the perturbations at the inner boundary but much more sensitive to perturbations at the outer boundary; In Fig.1 (e), the field is mainly distributed close to the inner boundary in the cloak with a concave transformation function $f_3(r)$. This so called concave-transformed cloak is not sensitive to the perturbations at the outer boundary, but it is sensitive to tiny perturbations at the inner boundary. In a word, the field inside the cloak is larger in the position where the differential of the function $f(r)$ is larger. Thus by choosing a function like $f_4(r)$, the differential of which is zero at both $R_1$ and $R_2$, we can get a cloak in which the field is mainly distributed near the central region of the coating and approaches to zero at both boundaries. In fact, if we choose a transformation function $f(r)$ that satisfies $f'(R_1)=0$ and $f'(R_2)=1$, the cloak will be insensitive to perturbations at neither the outer boundary nor the inner boundary, however, it will be sensitive to the perturbations in the central region of the cloak.

## IV. CONCENTRATORS

From the discussion of Part III, we can see if $f(r)$ is continuous at the boundary of $R_1$ and $R_2$, the scattered field is equal to zero, the fields in the whole region will still take the forms of equation (13). For any medium that satisfies $\frac{\varepsilon}{\varepsilon_0}=\frac{\mu}{\mu_0}=a$ (corresponds to $f(r)=ar$), we can cover it with a certain coating to make it invisible to the detector outside, but compared with the cloak case where no energy can be transmitted inside, this coating is different in that energy can still penetrate into the core. Fig.2 (a) shows four different cases with four different cores (a = 0.5, 1.5, 2, 2.5) and their coated layers created with four different transformation functions, $f_1(r)$, $f_2(r)$, $f_3(r)$, and $f_4(r)$, respectively. For simplicity, we choose four linear functions determined by $a$, which can be see in Fig.2 (a). $R_2$ is set to be 0.2 m and $R_1$ is equal to 0.1 m. Fig.2 (b) shows the corresponding constitutive parameter components of different coatings. Using the aforementioned formulations, we can calculate the field solutions for these four cases. The $E_x$ field distributions of these four specific cases under an $E_x$ polarized plane wave incidence along z direction are displayed in Fig2. (c), (d), (e), and (f), respectively.

Since the vector potential takes exact the same form as equation (10), the field distributions shown in Fig.2 (c-f) are determined by the relative parameter $a$:

When $0 < a < 1$, the power transmitted through the core is relative small, most power transmitted through the coating layer. An example of this case is shown in Fig.2(c), where $a = 0.5$. When $a = 0$, the coating will reduce to a perfect cloak.

When $a \geq 1$, the coating can be treated as a concentrator. When $1 < a < R_2/R_1$ (here $R_2/R_1 = 2$), most power will transmit through the coating, an example of this case is shown in Fig.2 (d), where $a = 1.5$.

While at $a = R_2/R_1$, as an extreme case with $f'(r) = 0$ in the coating layer, all the power that transmit into (or out from) the inner region is along the radii of the coating, as shown in Fig. 2 (e). In this case, the radial components of $\varepsilon$ and $\mu$ tend to infinite while the tangential components equal zero. As we know, when a wave is incident from vacuum to a medium, at the boundary of which the tangential $\varepsilon$ ($\mu$) is infinite or the radial $\varepsilon$ ($\mu$) is zero, it will be totally reflected due to the surface current or surface voltage at the boundary [16]. Our case is different in that $k_t = \omega\sqrt{\varepsilon_r \mu_t} = k_0 \frac{f(r)}{r}$ has a finite value, which means the wave number is finite, so the electromagnetic wave can still propagate into this media. With equations 13, we can see $E_r = 0$ and $H_r = 0$, showing that the Poynting power is always along the radii of the sphere. Similarly, with equations (9) and equations (13) we can find that the non-zero components of $D$ and $B$ are always along the radial direction, which means the wave vector of the electromagnetic wave is always perpendicular to its Poynting power.

A more interesting case is when $a > R_2/R_1$, the permittivity and permeability of the coating is negative. The power that flows through the inner region (core) is larger than the power flows through the whole concentrator, because there is always some energy circulating between the coating and the inner media, as shown in Fig. 2 (f). The proportion of the power flow through the inner region $P_{in}$ to the power flow through the concentrator $P_{out}$ is:

$$P_{in}/P_{out} = \frac{\int_0^{2\pi} d\varphi \int_0^{\frac{\pi}{2}} \hat{n}_1 \cdot (\bar{E} \times \bar{H}) r^2 \sin\theta d\theta \Big|_{r=R_1}}{\int_0^{2\pi} d\varphi \int_0^{\frac{\pi}{2}} \hat{n}_2 \cdot (\bar{E} \times \bar{H}) r^2 \sin\theta d\theta \Big|_{r=R_2}} = a^2 \left(R_1/R_2\right)^2, \qquad (14)$$

where $\hat{n}_1$ represents the surface normal of the inner boundary $R_1$, $\hat{n}_2$ represents surface normal of the outer boundary $R_2$. As shown in Fig. 2 (f), there is no scattering, but the power is magnified inside ($P_{in} > P_{out}$). The reason of this interesting phenomenon is that the case we consider here is in the time harmonic state. Before reaching this steady time harmonic state, there is scattered field outside, and the energy is getting stored in the coating. When the steady state is reached, the stored energy will circulate between the coating and inner media. Larger $a$ can lead to more energy stored in the concentrator.

## V. FIELD ROTATOR

Cylindrical rotation coating was first proposed by H. Chen and C. T. Chan [5]. In this session we will propose a three-dimensional (spherical) rotation coating, and calculate all the components of the fields through our method. The analytical results show similar behaviors with the simulation results in Ref [5] as the electromagnetic wave is propagating in the x-y plane. Furthermore, we demonstrate if the incident wave is not in the x-y plane, not only the wave front of the wave but also the polarization of the fields will be rotated.

Consider the following transformation: $r' = r$, $\theta' = \theta$, and $\varphi' = \varphi + g(r)$, where $g(R_2) = 0$ and $g(R_1) = \varphi_0$. Using equation (1) the permittivity and permeability tensor components of the coating shell can be given as:

$$\bar{\bar{\varepsilon}} = \begin{bmatrix} 1 & 0 & -\alpha \\ 0 & 1 & 0 \\ -\alpha & 0 & 1+\alpha^2 \end{bmatrix} \varepsilon_0, \qquad \bar{\bar{\mu}} = \mu_0 \begin{bmatrix} 1 & 0 & -\alpha \\ 0 & 1 & 0 \\ -\alpha & 0 & 1+\alpha^2 \end{bmatrix}, \qquad (15)$$

where $\alpha = r \sin\theta g'(r)$. The above parameters are all expressed in spherical coordinate. With this relation, the field is matched at both the outer boundary $R_2$ and the inner boundary $R_1$, but the tangential angle $\varphi$ has been rotated by an angle $\varphi_0$ from outer boundary to inner boundary of the coating. So the wave propagating in x-y plane will change its direction by $\varphi_0$ inside the enclosed domain with respect to that outside the coating. Consider a plane wave incident upon the coated sphere along the x direction with unit amplitude of electric field $\bar{E}_i = \hat{y} e^{ik_0 x}$, since $r = R_2$ and $r = R_1$ are both matched boundaries, with equation (7), the vector $\bar{A}_{TM}$ and $\bar{A}_{TE}$ inside of the coating can be expressed as

$$\bar{A}_{TM} = \hat{r} \frac{1}{\omega} \sum_{m,n} \psi_n(k_0 r) \left[ a_{m,n}^{TM} T_{mn}^e (\varphi + g(r)) + b_{m,n}^{TM} T_{mn}^o (\varphi + g(r)) \right]$$

$$\bar{A}_{TE} = \hat{r} \frac{1}{\omega \eta_0} \sum_{m,n} \psi_n(k_0 r) \left[ a_{m,n}^{TE} T_{mn}^e (\varphi + g(r)) + b_{m,n}^{TE} T_{mn}^o (\varphi + g(r)) \right]$$

$$(16)$$

where $T_{mn}^e(\theta,\varphi) = P_n^m(\cos\theta)\cos(m\varphi)$, $T_{mn}^o(\theta,\varphi) = P_n^m(\cos\theta)\sin(m\varphi)$, $a_{m,n}^{TM}$, $b_{m,n}^{TM}$ and $a_{m,n}^{TE}$, $b_{m,n}^{TE}$ are the spherical expansion coefficients of $e^{ik_0 \sin\theta\cos\varphi} \sin\theta\sin\varphi$ and

$e^{ik_0 \sin\theta \cos\varphi} \cos\theta$ respectively. substituting equations (16) into equations (8), the field in the coating can be written in the closed form:

$$E_r = \sin\theta \left[ \sin(\varphi + g(r)) + rg'(r)\cos(\varphi + g(r)) \right] e^{ik_0 \sin\theta \cos(\varphi + g(r))} \qquad (17\text{-}1)$$

$$E_\theta = \cos\theta \sin(\varphi + g(r)) e^{ik_0 \sin\theta \cos(\varphi + g(r))} \qquad (17\text{-}2)$$

$$E_\varphi = \cos(\varphi + g(r)) e^{ik_0 \sin\theta \cos(\varphi + g(r))} \qquad (17\text{-}3)$$

$$H_r = \frac{\cos\theta}{\eta_0} e^{ik_0 \sin\theta \cos(\varphi + g(r))} \qquad (17\text{-}4)$$

$$H_\theta = -\frac{\sin\theta}{\eta_0} e^{ik_0 \sin\theta \cos(\varphi + g(r))} \qquad (17\text{-}5)$$

$$H_\varphi = 0 \qquad (17\text{-}6)$$

If we choose the following transformation: $g(r) = \varphi_0 \dfrac{\ln r - \ln R_2}{\ln R_1 - \ln R_2}$, the above calculation can be simplified [5]. As a result all the component of $\bar{\bar{\varepsilon}}$ and $\bar{\bar{\mu}}$ are independent of $r$. Fig.3 (a) and (b) show the $H_z$ distribution for $\varphi_0 = \dfrac{\pi}{2}$ and $\varphi_0 = \pi$ respectively.

If the wave with unit electric field $\bar{E}_i = \hat{x} e^{ik_0 z}$ is incident along z axis (perpendicular to the x-y plane) onto the rotator with $\varphi_0 = \dfrac{\pi}{2}$, following the same steps, we can get the distribution of $E_x$ and $E_y$, as shown in Fig. 3 (c) and (d), respectively. It is interesting to see that in this case, instead of the wave front of the electromagnetic wave, it is the polarization of the fields that is rotated as the wave passing into the coating. Therefore, this kind of spherical field rotator can function as a wave vector rotator as well as a polarization rotator. Tuning the orientation of the spherical rotator with respect to the incident wave direction can control the functionality of the spherical rotators.

## VI. CONCLUSION

In this paper, we summarized a generalized formulation on how to use the transformations to obtain different devices by combining the merits of both coordinate transformation and the Mie scattering solutions. We show by mathematical analysis that the coordinate transformation to the Maxwell equations can be in a generalized form: any continuous functions can be adopted in the transformation, and different type of functions will bring different characteristics of the EM behaviors in the transformed space. Starting from the formulation deduced in this paper, invisibility cloaks, concentrators, and rotators can be easily obtained by simply selecting different scalar transformation functions, and the internal field distributions in these devices can also be controlled by tuning the shapes of the functions. Various examples for the design of cloaks, concentrators, and field rotator are given to demonstrate the validity of the formulation and the very high computational efficiency. Our paper presents a very useful tool in the analysis and design for the invisibility cloak, EM concentrators, field rotators, and similar novel devices.

**ACKNOWLEDGEMENT**

This work is sponsored by the Chinese National Science Foundation under Grant Nos. 60531020 and 60671003, the NCET-07-0750, the China Postdoctoral Science Foundation under Grant No. 20060390331, the ONR under Contract No. N00014-01-1-0713, and the Department of the Air Force under Air Force Contract No. F19628-00-C-0002.

[18] H.C.van de Hulst, *Light Scattering by Small Particles,* Dover, New York, (1957).

**Figure Captions**

FIG. 1 (a) Schematic figure of the transformation functions $f(r)$ in four cases:
(Case I) $f_1(r) = \dfrac{R_2}{(R_2 - R_1)}(r - R_1)$, (Case II) $f_2(r) = \dfrac{R_2}{(R_2 - R_1)^2}(r - R_1)^2$ with $f_2'(R_1) = 0$,
(Case III) $f_3(r) = -\dfrac{R_2}{(R_2 - R_1)^2}(r - R_2)^2 + R_2$ with $f_3'(R_2) = 0$, and (Case IV) $f_4(r)$
with $f_4'(R_1) = 0$ and $f_4'(R_2) = 0$. (b) The permittivity and permeability components calculated from the corresponding four cases. (c), (d), (e), and (f) show the $E_x$ fields distribution and Poynting vectors due to $E_x$ polarized wave incidence onto a cloak created with the four different transformation functions $f_1(r)$, $f_2(r)$, $f_3(r)$, and $f_4(r)$, respectively.

FIG. 2 (a) Schematic figure of the four different configurations: four different cores ($a = 0.5, 1.5, 2, 2.5$) and their coated layers created with four transformation functions, $f_1(r)$, $f_2(r)$, $f_3(r)$, and $f_4(r)$, respectively. (b) The permittivity and permeability components of the coating in the four cases. The radial component of the constitutive parameter in the coating layer created with $f_3(r)$ is infinite, so it is not plotted out here. (c), (d), (e), and (f) are the calculated $E_x$ field distribution and Poynting vectors due to $E_x$ polarized wave incidence onto the coated sphere.

FIG. 3 (a) Distribution of $H_z$ and Poynting vectors as the wave is incident along x direction for $\varphi_0 = \dfrac{\pi}{2}$ (b) Distribution of $H_z$ and Poynting vectors as the wave is incident along x direction for $\varphi_0 = \pi$ (c) Distribution of $E_x$ as a $E_x$ polarized wave incident along z direction for $\varphi_0 = \frac{\pi}{2}$ (d) Distribution of $E_y$ as a $E_x$ polarized wave incident along z direction $\varphi_0 = \frac{\pi}{2}$.

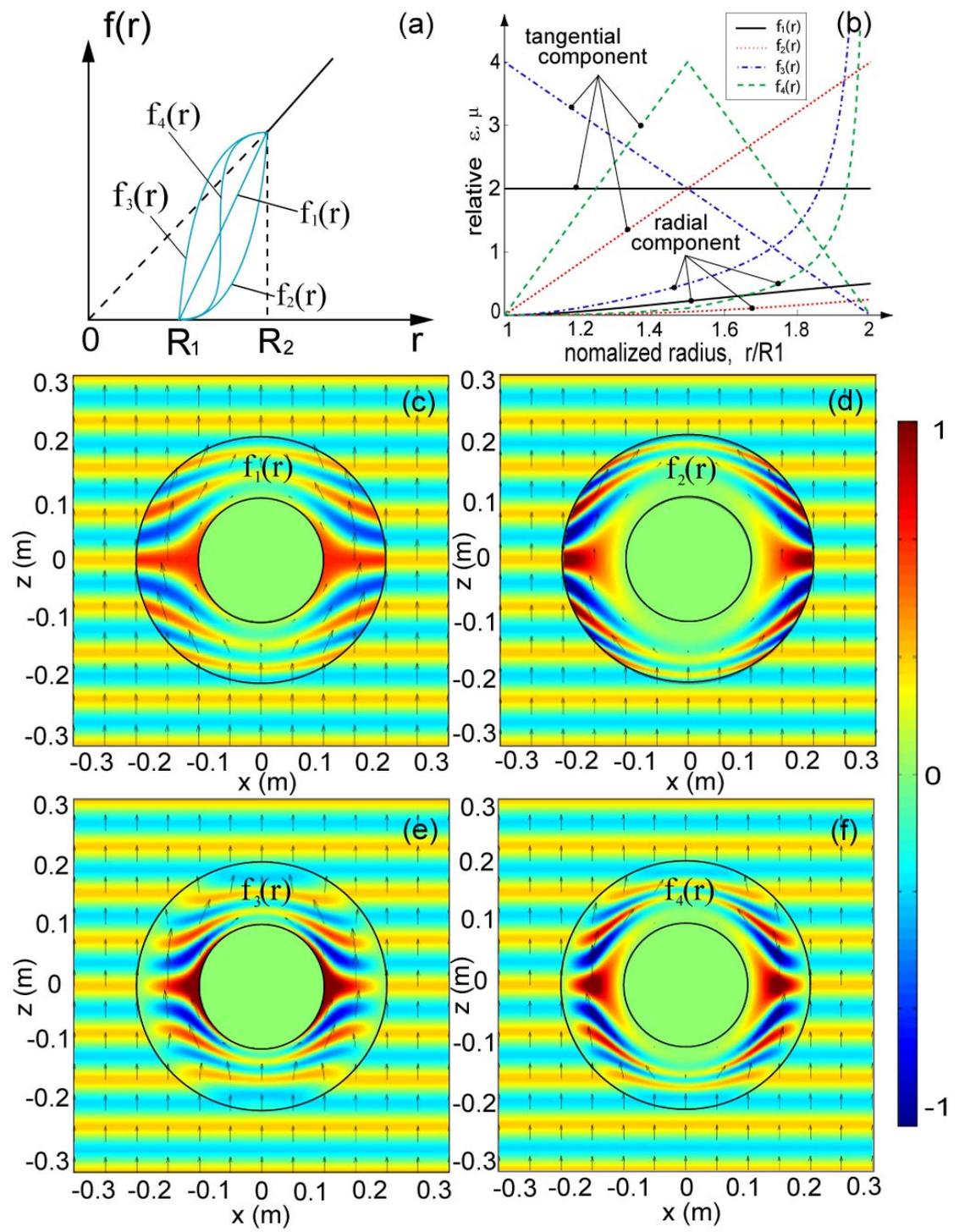

FIG. 2

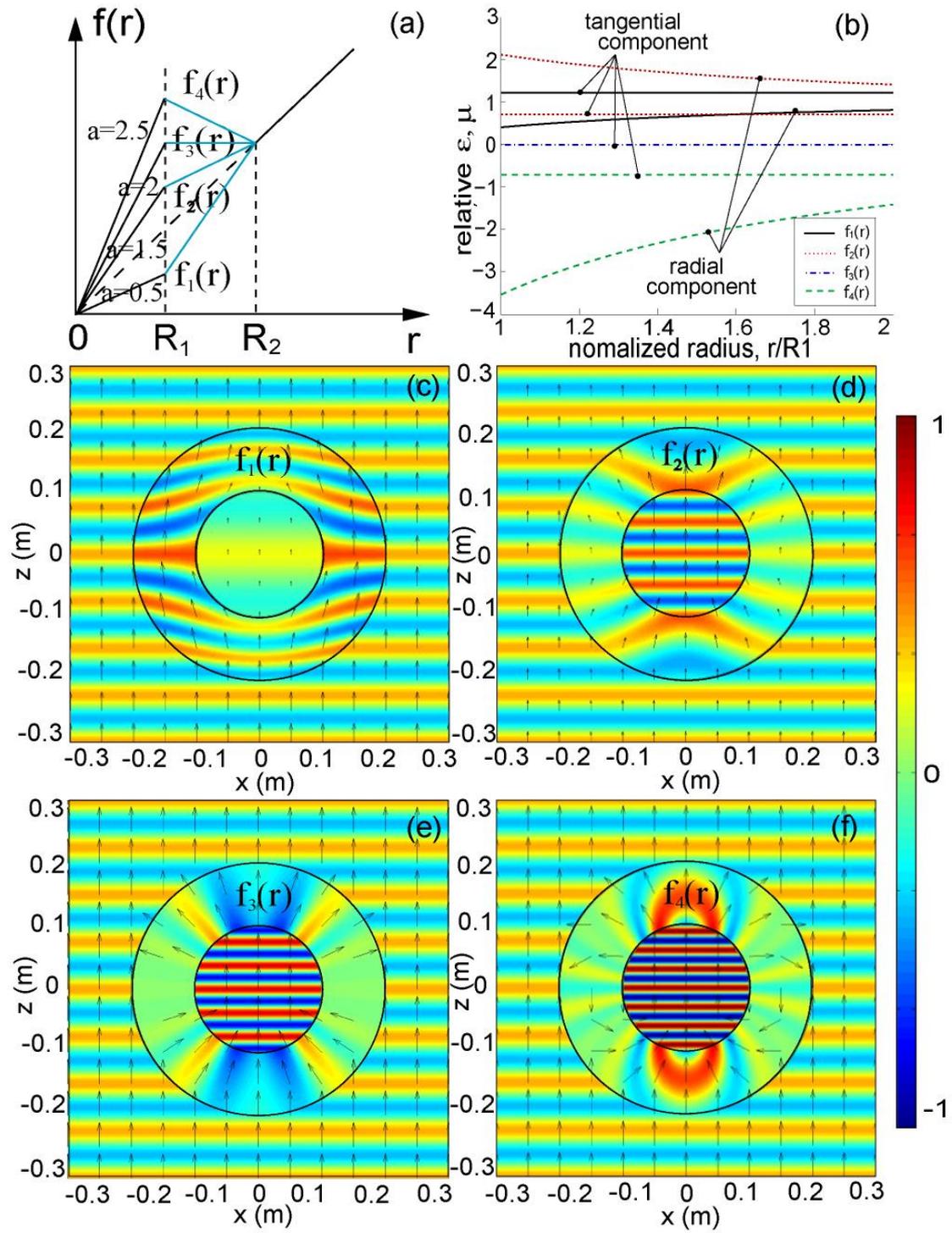

FIG 3

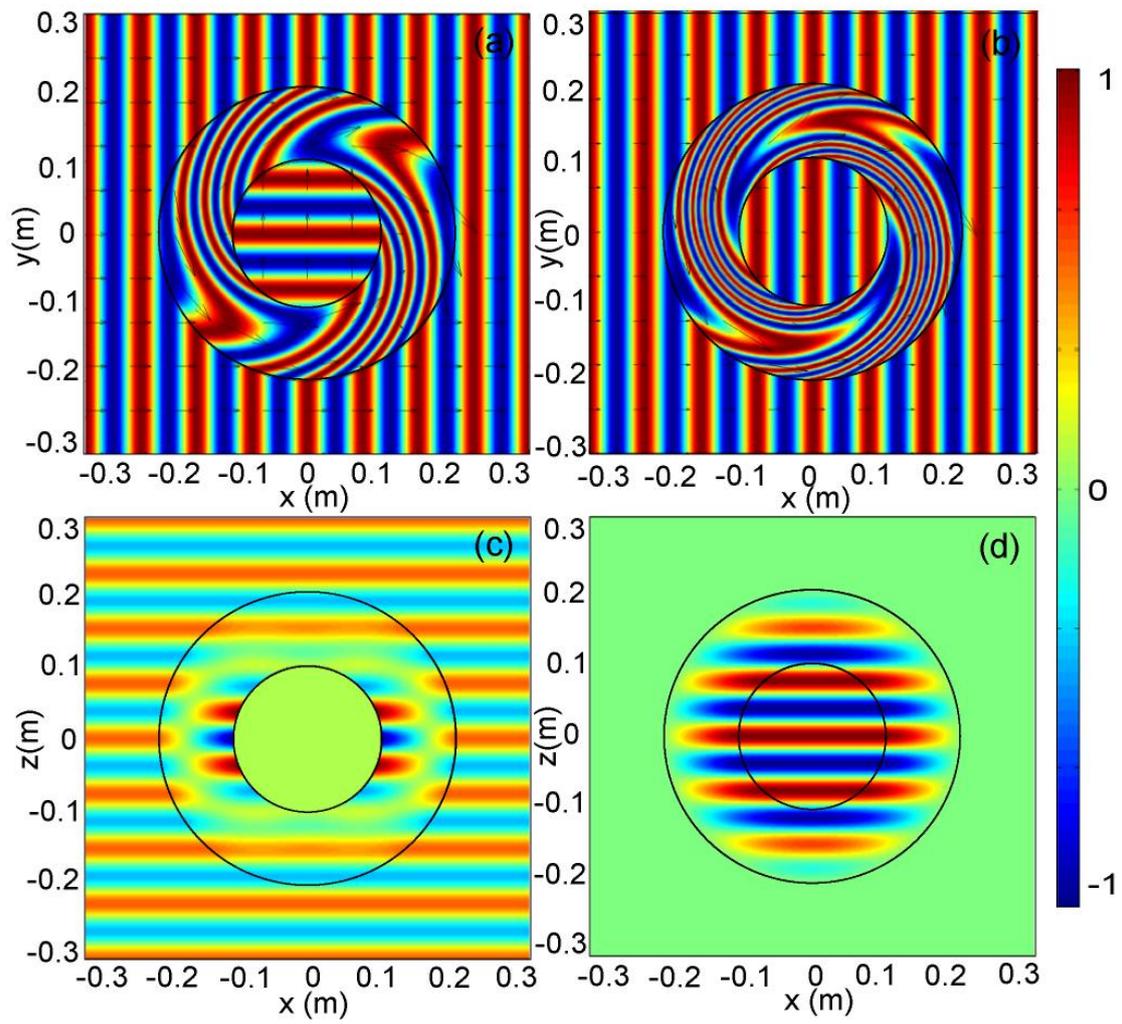